\documentclass[pre,preprint,amsmath]{revtex4-1}

\newcommand{\beq}{\begin{equation}}
\newcommand{\eeq}{\end{equation}}
\newcommand{\mf}{mean-field}
\newcommand{\lb}{\langle}
\newcommand{\rb}{\rangle}
\newcommand{\lam}{\lambda}
\newcommand{\wmf}{w_{\rm mf}}
\newcommand{\wpp}{\tilde w}
\newcommand{\fc}{fully connected}
\newcommand{\ys}{Yard-Sale}

\usepackage{graphicx}
\usepackage{verbatim}

\begin{document}

\title{Mean-field theory of an asset exchange model with economic growth and wealth distribution}

\author{W. Klein}
\email{klein@bu.edu}
\affiliation{Department of Physics, Boston University, Boston, Massachusetts 02215}
\affiliation{Center for Computational Science, Boston University, Boston, Massachusetts 02215}

\author{N. Lubbers}
\altaffiliation[Present address: ]{Los Alamos National Laboratory, Los Alamos, New Mexico 87545}
\affiliation{Department of Physics, Boston University, Boston, Massachusetts 02215}

\author{Kang K. L. Liu}

\affiliation{Department of Physics, Boston University, Boston, Massachusetts 02215}

\author{T. Khouw}
\affiliation{Department of Physics, Boston University, Boston, Massachusetts 02215}

\author{Harvey Gould}
\affiliation{Department of Physics, Boston University, Boston, Massachusetts 02215}
\affiliation{Department of Physics, Clark University, Worcester, Massachusetts 01610}

\begin{abstract}

We develop a mean-field theory of the growth, exchange and distribution (GED) model introduced by Kang et al.\ (preceding paper) that accurately describes the phase transition in the limit that the number of agents $N$ approaches infinity. 
The GED model is a generalization of the \ys\ model in which the additional wealth added by economic growth is nonuniformly distributed to the agents according to their wealth in a way determined by the parameter $\lam$. The model was shown numerically to have a phase transition at $\lam=1$ and be characterized by critical exponents and critical slowing down. Our mean-field treatment of the GED model correctly predicts the existence of the phase transition, critical slowing down, the values of the critical exponents, and introduces an energy whose probability satisfies the Boltzmann distribution for $\lam < 1$, implying that the system is in thermodynamic equilibrium in the limit that $N \to \infty$. 
We show that the values of the critical exponents obtained by varying $\lam$ for a fixed value of $N$ do not satisfy the usual scaling laws, but do satisfy scaling if a combination of parameters, which we refer to as the Ginzburg parameter, is much greater than one and is held constant. We discuss possible implications of our results for understanding economic systems and the subtle nature of the mean-field limit in systems with both additive and multiplicative noise.
\end{abstract}

\maketitle

\section{Introduction}

Agent-based asset exchange models have become useful~\cite{aem1, aem2, krapiv, aem3, aem4, aem5, oneofthefirst, ijmp, hayes, Burda, Chakraborti2} for studying the effects of chance on the distribution of wealth. These models consist of $N$ agents that can exchange wealth through pairwise encounters. 
Examples of the exchange mechanism include the transfer of a fixed amount of wealth and the exchange of a fixed percentage of the average of the wealth of the two agents. The common feature of these models is that the winner in the exchange is determined by chance.

Of particular interest is the Yard-Sale model~\cite{aem2, oneofthefirst, hayes, popbogo, ijmp, Burda, bogo, bogo2, bogo3, devitt, probyardsale, Moukarzel, rosser} in which pairs of agents are chosen at random and one is designated as the winner with a probability usually taken to be 1/2. The winner receives a fraction $f$ of the wealth of the poorer agent. The result is that after many exchanges, one agent gains almost all of the wealth, a phenomena known as wealth condensation.

In this paper we study a generalization of the Yard-Sale model~\cite{kang} in which a fixed percentage $\mu$ of the total wealth is added to the system after $N$ exchanges.
The added wealth is distributed according to
\beq
\Delta w_{i}(t) = \mu W(t)\frac{w^{\lam}_{i}(t) }{ \sum_{j=1}^{N}w^{\lam}_{j}(t)}, \label{DW}
\eeq
where $w_{i}(t)$ is the wealth of agent $i$ at time $t$ and $\lam \geq 0$ is the distribution parameter. The quantity $\mu W(t)$ is the change in the total wealth in the system at time $t$ due to economic growth, where $W(t)$ is the total wealth of the system at time $t$, and the parameter $\mu$ is the rate of growth.
One unit of time corresponds to $N^{2}$ exchanges. 
This distribution mechanism is justified by economic data in 
the Appendix of Ref.~\onlinecite{kang}.

This model, which we denote as the growth, exchange and distribution (GED) model, was investigated numerically~\cite{kang} and shown to have a phase transition at $\lam = 1$. For $0 < \lam < 1$ the wealth is not distributed uniformly, but wealth condensation is avoided. As $\lam$ approaches 1 from below, the wealth distribution becomes more skewed toward the rich. However, there is economic mobility and poorer agents can become richer and richer agents can become poorer. In addition, every agent's wealth increases exponentially as $e^{\mu t}$ due to economic growth as the system evolves with time. In contrast, for $\lam \geq 1$ there is wealth condensation as was found in the original Yard-Sale model ($\mu=0$), and there is no economic mobility.

Numerical investigations indicate that the phase transition at $\lam = 1$ is continuous~\cite{kang}. The order parameter $\phi$ is defined as the fraction of the wealth held by all of the agents except the richest and goes to zero as $\lam \rightarrow 1^-$ (for $N\rightarrow \infty$).
Three exponents were introduced in Ref.~\onlinecite{kang} to characterize the behavior of various quantities as $\lam \to 1^-$, including the order parameter $\phi \sim (1-\lam)^\beta$ and the susceptibility $\chi \sim (1-\lam)^{-\gamma}$ (the variance of the order parameter). 
As we will discuss in Sec.~\ref{sec:gedenergy}, we can define the total energy of the system and introduce the exponent $\alpha$ to characterize the critical behavior of the nonanalytic part of the mean energy as $(1 - \lam)^{1 - \alpha}$. Similarly, we relate the specific heat to the variance of the energy and characterize its divergence as $(1-\lam)^{-\alpha}$. 
Because there is no length scale in the GED model, there is no obvious way of defining a correlation length exponent.

Simulations at fixed values of $N$ in Ref.~\cite{kang} yield the estimates
$\beta \approx 0$, $\gamma \approx 1$, $1 - \alpha \approx -1$ and $\alpha\approx 2$. These values do not satisfy the scaling law~\cite{scalingref}
\beq
\alpha + 2\beta + \gamma = 2, \label{scalinglaw}
\eeq
and do not appear to correspond to any known universality class.
 
In this paper we present a mean-field treatment of the GED model and find that the interpretation of the critical exponents is subtle. The theory shows that if the critical behavior is interpreted correctly, the exponents do satisfy Eq.~\eqref{scalinglaw} with $\beta=0$, $\gamma=1$ and $\alpha = 1$. Moreover, we can define an energy and a Hamiltonian that allows us to obtain an equilibrium (Boltzmann) description of the GED model in the limit that the number of agents $N \to \infty$. The mean-field theory results are consistent with the simulations~\cite{kang}.

In addition to casting light on the nature of the critical point, the mean-field approach predicts that for $\lam < 1$, the wealth distribution can be made less skewed toward the rich by increased growth for fixed $N$, $\lam$, and $f$. The mean-field approach also indicates that wealth inequality can be reduced for 
fixed $\lam < 1$ and fixed $N$ and $\mu$ by decreasing the value of $f$, corresponding to decreasing the magnitude of the noise. 
However, for $\lam \geq 1$, economic growth does not avoid wealth condensation, and there is no economic mobility.

The structure of the remainder of the paper is as follows. In Sec.~\ref{sec:equation} we construct an exact differential equation for the GED model and then introduce the mean-field approximation to the equation. In Sec.~\ref{sec:transition} we show that there is a phase transition at $\lam = 1$ with critical slowing down, and obtain the values of the critical exponents $\beta$ and $\gamma$. In Sec.~\ref{sec:gedenergy} we introduce the Ginzburg parameter, define the total energy of the system, and determine the critical exponent $\alpha$. In Sec.~\ref{sec:comparison} we compare the predicted mean-field exponents with the numerical estimates. 
We discuss the role of multiplicative noise in the GED model in Sec.~\ref{sec:multnoise}
and examine the relation between the GED model and the geometric random walk in Sec.~\ref{sec:geometric}. Finally in Sec.~\ref{sec:discussion}, we discuss the implication of these results for critical phenomena in fully connected systems and systems with long but finite-range interactions, and discuss the implication of our results for the study of economic systems. Because the GED model is similar in several ways to the fully connected Ising model, we review some aspects of that model in the Appendix and discuss the Ginzburg criterion as a self-consistency check on the applicability of mean-field theory.

\section{\label{sec:equation}Exact and Mean-Field Equations}

The rate of change of the wealth of agent $i$ is given by a formally exact stochastic difference equation
\begin{eqnarray}
\frac{\Delta w_{i}(t)}{1} &=& f \sum_{j} \Theta\big[w_{i}(t) - w_{j}(t)\big]\eta_{ij}(t) w_{j}(t) \nonumber \\
&& {}+ f \sum_{j}\Big\{1 - \Theta \big [w_{i}(t) - w_{j}(t)\big] \Big \} \eta_{ij(t)}w_{i}(t) +
\mu W(t) \frac{w_{i}(t)^{\lam}}{S(t)}. \label{exacteq}
\end{eqnarray}
The denominator on the left-hand side of Eq.~\eqref{exacteq} is written as 1 to emphasize that Eq.~\eqref{exacteq} is a difference equation rather than a differential equation.
Here
\beq
\Theta \big (w_{i} - w_{j}\big) =
\begin{cases}
1 & \big(w_{i} \geq w_{j}\big)\\
0 & \big(w_{i} < w_{j}\big),
\end{cases}
\eeq
and
\beq
S(t) = \sum_{i} w^{\lam}_{i}(t).\label{S}
\eeq
The parameter $f$ is the fraction of the poorer agents's wealth that is exchanged, $\mu$ is the fraction of the total wealth that is added after $N$ exchanges, the parameter $\lambda$ determines the distribution of the added economic growth, and $\eta_{ij}(t)$ for $\i \neq j$ is a time-dependent random matrix element such that
\beq
\eta_{ij}(t) = 
\begin{cases}
0 & \mbox{agents $i$ and $j$ do not exchange wealth} \\
1 & \mbox{wealth is transferred from agent $j$ to agent $i$}\\
-1 & \mbox{wealth is transferred from agent $i$ to agent $j$}.
\end{cases}
\eeq
($\eta_{ij} = 0$ if $i = j$.) The matrix elements of $\eta$ can be chosen from any probability distribution with the constraint that if $\eta_{ij} = \pm 1$, then $\eta_{ji} =\mp 1$. This condition imposes the constraint that the exchange conserves the total wealth. 

To obtain a differential equation we multiply and divide the denominator on the left-hand side of Eq.~\eqref{exacteq}
by $N$, the number of agents. Because we will take the limit $N\rightarrow \infty$ and take 
one time unit to correspond to $N^{2}$ exchanges, we have that $1/N\rightarrow dt$. 
Note that in the simulations of Ref.~\cite{kang}, $N$ exchanges was chosen as the unit of time. In this case each agent will, on the average, exchange wealth with only one other agent and hence one exchange described by the difference equation would not take place in an infinitesimal amount of time. One exchange per agent does take place in an infinitesimal time if one time unit corresponds to $N^{2}$ exchanges during which each agent exchanges wealth with every other agent on the average.

The parameters $f$ and $\mu$ in Eq.~\eqref{exacteq} are the rates of exchange and growth, respectively, and are defined per $N$ exchanges to be consistent with the simulations. 
To obtain a consistent differential equation, these rates need to be scaled by $N$. 
We let
\beq
f = f_0/N \mbox{ and } \mu = \mu_0/N, \label{eq:scaling}
\eeq
and assume that $f_0$ and $\mu_0$ are independent of $N$. We will see that these theoretical considerations imply that the values of the
parameters $f$ and $\mu$ chosen in the simulations must be scaled with $N$ if the Ginzburg parameter is held fixed. 

Because wealth is added to the system after every $N$ exchanges, the total wealth in the system at time $t$ is given by
\beq
W(t) = W(0)e^{\mu_0 t}. \label{M}
\eeq

With these considerations Eq.~\eqref{exacteq} becomes
\begin{align}
\frac{dw_{i}(t)}{dt} &= {f_0} \sum_{j} \Theta\big[w_{i}(t) - w_{j}(t)\big]\eta_{ij}(t) w_{j}(t) \nonumber \\
&{}+ {f_0}\sum_{j}\Big\{1 - \Theta \big [w_{i}(t) - w_{j}(t)\big] \Big \} \eta_{ij}(t)w_{i}(t) +
{\mu_0} W(t) \frac{w_{i}(t)^{\lam}}{S(t)}. \label{exacteq2}
\end{align}

To obtain a mean-field theory, we choose an agent whose wealth is $w(t)$ and let $\wmf(t)$ be the mean wealth of the remaining agents. That is,
\beq
\wmf(t) = \frac{W(t) - w(t)}{N - 1}. \label{avwealth}
\eeq
The mean-field version of Eq.~\eqref{exacteq2} is
\begin{eqnarray}
\frac{dw(t)}{dt} & =& f_0 \Theta\big [w(t) - \wmf(t)\big]\eta \wmf(t) + f_0 \Big [1 - \Theta \big [w(t) - \wmf(t)\big] \Big] \eta w(t) \nonumber \\
&&{} +
\mu_0 W(t) \frac{w(t)^{\lam}}{S(t)}. \label{mfeq1}
\end{eqnarray}
The quantity $S(t)$ defined in Eq.~\eqref{S} becomes
\beq
S(t) = w^\lam(t) + (N-1) \wmf^\lam(t). \label{kangsum}
\eeq

To obtain a mean-field description we have effectively coarse grained the exchanges between the chosen agent and the remaining $N - 1$ agents in time, which implies a coarse graining of the noise associated with the coin flips that determine the exchange of wealth. By using the central limit theorem, we can take the noise in Eq.~\eqref{mfeq1} to be random Gaussian. 
This assumption would not be valid if the chosen agent interacted with only one other agent in one unit of time. 
However, because the unit of time corresponds to $N^{2}$ exchanges, the chosen agent interacts with $N - 1$ 
other agents and coarse graining in time makes sense. The coarse graining of the noise is another reason why it is necessary to choose $N^{2}$ exchanges to be one unit of time in the mean-field theory.

It will be convenient to write the growth term in Eq.~\eqref{mfeq1} as
\beq
\mu_0 W(t)\frac{w^{\lam}(t)}{S(t)} = \mu_0 W(t) \frac{[w(t)/W(t)]^{\lam}}{[w(t)/W(t)]^{\lam} + (N-1)^{1-\lam}[1 - w(t)/W(t)]^{\lam}}, \label{growth3}
\eeq
where we have used Eqs.~\eqref{avwealth} and \eqref{kangsum} and divided the numerator and denominator by $W^{\lam}(t)$. 

To simplify Eq.~\eqref{mfeq1}, we first assume that $w(t) < \wmf(t)$; that is, the wealth of the chosen agent is less than the mean wealth of the remaining $N-1$ agents. We use Eqs.~\eqref{exacteq2} and \eqref{growth3} to obtain
\beq
\frac{dw(t)}{dt} = f_0\eta(t) w(t) + \mu_0 W(t) \frac{[w(t)/W(t)]^{\lam} }{[w(t)/W(t)]^{\lam} + (N-1)^{1-\lam}[1 - w(t)/W(t)]^{\lam}}. \label{mfeq}
\eeq

We divide both sides of Eq.~\eqref{mfeq} by $W(t)$ and rewrite
Eq.~\eqref{mfeq} as
\beq
\frac{d}{dt}\Big(\frac{w(t)}{W(t)}\Big) = f_0 \eta(t) \frac{w(t)}{W(t)} + \mu_0 \frac{[w(t)/W(t)]^{\lam}}{ [w(t)/W(t)]^{\lam} + (N-1)^{1-\lam}[1 - w(t)/W(t)]^{\lam}} - \mu_0 \frac{w(t)}{W(t)}, \label{mfeq2}
\eeq
where we have used the relation [see Eq.~\eqref{M}]
\beq
\frac{1}{W(t)}\frac{dw(t)}{dt} = \frac{d}{dt}\Big (\frac{w(t)}{W(t)}\Big) + \mu_0\frac{w(t)}{W(t)}.\label{diffeq}
\eeq

We next introduce the scaled wealth fraction
\beq
x(t) \equiv \frac{w(t)}{W(t)}, \label{ydef}
\eeq
and rewrite Eq.~\eqref{mfeq2} as
\begin{subequations}
\label{poorer}
\begin{align}
\frac{dx(t)}{dt} & = R(x, \eta, t) \\
\noalign{\noindent with}
R(x, \eta, t) & \equiv f_0 \eta(t) x(t) + \mu_0 \frac{x(t)^{\lam}}{x(t)^{\lam} + (N - 1)^{1-\lam}[1 - x(t)]^{\lam}} - \mu_0 x(t). \label{eq:R}
\end{align}
\end{subequations}
Equation~\eqref{poorer} expresses the time-dependence of the wealth of the chosen agent in contact with a mean-field representing the mean wealth of the remaining agents. Hence, the wealth of the chosen agent is not conserved. 

For $\mu_0 = 0$, the total wealth $W$
is a constant because the noise $\eta(t)$ that determines the wealth transfer from the mean-field wealth to the chosen agent is the negative of the noise that governs the wealth transfer from the chosen agent to the mean field.

It is easy to show that for zero noise, $R(x, \eta=0, t)=0$ for $x = 0$, 1, and $1/N$, and that these are the only fixed points of Eq.~\eqref{poorer} for $\lam \neq 1$. To determine the stability of the fixed points, we calculate the derivative $dR(x, \eta=0, t)/dx$ and obtain
\beq
\frac{dR(x, 0, t)}{dx} = \mu_0 \frac{\lam x^{\lam - 1} }{ x^{\lam} + (N - 1)^{1 - \lam}(1 - x)^{\lam}} - \mu_0 \frac{x^{\lam}[\lam x^{\lam - 1} - \lam(N - 1)^{1- \lam}(1 - x)^{\lam - 1}] } { [x^{\lam} + (N - 1)^{1 - \lam} (1 - x)^{\lam}]^{2}} - \mu_0, \label{deriv-fixed}
\eeq
where $x \equiv x(t)$. For $\lam < 1$, the derivatives at $x = 0$ and $x = 1$ are equal to $\infty$, which implies that these fixed points are unstable. The derivative at $x=1/N$ is equal to $\mu_0(\lam - 1)$, and hence the fixed point at $x =1/N$ is stable for $\lam< 1$. For $\lam > 1$ the derivative at $x = 1/N$ is positive so that this fixed point is unstable. The derivative at $x = 0$ and $x=1$ equals $-1$, and hence these fixed points are stable.

We next return to Eq.~\eqref{exacteq2} and consider the case for which $w(t) > \wmf(t)$. The growth term is the same as before. The exchange term in Eq.~\eqref{exacteq},
$
f_0 \sum_{j} \Theta\big (w_{i} - w_{j}\big )\eta_{ij}w_{j}
$,
becomes $f_0 \eta \wmf$. We use Eq.~\eqref{avwealth} to write
\beq
\frac{dw(t)}{dt} = f_0\eta(t) \frac{W(t) - w(t)} { N - 1} + \mu_0 W(t) \frac{[w(t)/W(t)]^{\lam} }{ [w(t)/W(t)]^{\lam} + (N-1)^{1-\lam}[1 - w(t)/W(t)]^{\lam}}.\label{deriv}
\eeq
From Eq.~\eqref{diffeq} and the definition of $x(t)$ in Eq.~\eqref{ydef} we have
\beq
\frac{dx(t)}{ dt} = f_0 \eta(t) \frac{1 - x(t)}{N - 1} + \mu_0 \frac{x(t)^{\lam}}{x(t)^{\lam} + (N - 1)^{1 - \lam}[1 - x(t)]^{\lam}} - \mu_0 x(t).\label{richer}
\eeq
Equation~\eqref{poorer} for the poorer agent and Eq.~\eqref{richer} for the richer agent are the same except for the noise term, and hence the fixed points are the same. We again use Eq.~\eqref{avwealth} to rewrite Eq.~\eqref{richer} as
\beq
\frac{dx(t)}{ dt} = f_0 \eta(t){x}_{\rm mf}(t) + \mu_0 \frac{x(t)^{\lam}}{x(t)^{\lam} + (N - 1)^{1 - \lam}[1 - x(t)]^{\lam}} - \mu_0 x(t),\label{richer2}
\eeq
where ${x}_{\rm mf}(t) = w_{\rm mf}(t)/W(t)$ is the fraction of the mean field agent's rescaled wealth. 
Note that $x(t)$ is of order $1/N$ as is ${x}_{\rm mf}(t)$. Equation~\eqref{richer2} will be used in Sec.~\ref{sec:transition} to discuss the phase transition and the critical exponents.

In summary, the fixed points for all values of $\lam$ are $x = 0$, 1, and $1/N$ for the \mf\ equations describing the wealth 
evolution of either the richer or poorer agent. For $\lam<1$, the fixed points at 0 and 1 are unstable, and the fixed point at $1/N$ is stable, corresponding to all agents having an equal share of the total wealth on average. For $\lam> 1$, the fixed points at 0 and 1 are stable, and the fixed point at $x = 1/N$ is unstable, which implies that if all the agents are assigned an equal amount of wealth at $t=0$, one agent will eventually accumulate all the wealth in a simulation of the model.
Note that if we use the equation for which the chosen agent is richer than the ``mean field'' agent, then the stable 
fixed point reached when $\lambda > 1$ is $x = 1$; similarly, if we chose the equation for which the chosen agent is poorer than the mean field agent, the stable fixed point reached for $\lambda > 1$ is $x = 0$.

\section{\label{sec:transition}The Phase Transition}

To analyze the phase transition at $\lam = 1$, we investigate Eq.~\eqref{richer2}, the mean-field differential equation for the richer agent, for $x\sim 1/N$ and $\lam$ close to $1^-$.
We let
\beq
x(t) = \frac{1}{N} - \delta(t), \label{eq:delta}
\eeq
assume $N \delta \ll 1$, and expand the second term on the right-hand side of Eq.~\eqref{eq:R} to first order in $N \delta$. After some straightforward algebra we find that
\beq
\frac{d\delta(t)}{dt} = f_0 \eta(t)x_{\rm mf}(t) - \mu_0 (1 - \lam) \delta(t), \label{fract4}
\eeq
We multiply both sides of Eq.~\eqref{fract4} by $N$ to obtain
\beq
\frac{dN\delta(t)}{dt} = f_0 \eta(t){\tilde w}_{\rm mf} - \mu_0 (1 - \lam) N\delta(t), \label{fract5}
\eeq
where ${\tilde w}_{\rm mf} = Nx_{\rm mf}$. We write ${\tilde w}_{\rm mf} = 1 - N \delta$, let
\beq
\phi = N \delta, \label{eq:orderparameter}
\eeq
and rewrite Eq.~\eqref{fract5} as
\beq
\frac{d\phi(t)}{dt} = f_0 \eta(t) \big [1 - \phi(t)\big] - \mu_0 (1 - \lam)\phi(t). \label{fract51}
\eeq

As mentioned, we can assume the noise $\eta(t)$ to be associated with a random Gaussian distribution of coin flips. Note that $\eta$ is the average over $N$ coin flips and hence should scale as ${\sqrt N}/N \sim 1/{\sqrt N}$. 
Hence $\eta(t)$ in Eq.~\eqref{fract51} is order $1/\sqrt N$, which implies that $\phi(t)\sim 1/\sqrt N$ and justifies our neglect of terms higher than first order. Simulations in Ref.~\cite{kang} show that the fluctuations are dominated by those near the $1/N$ fixed point.

Because $\phi(t) \sim 1/\sqrt N \ll 1$ for $N \gg 1$, we can ignore $\phi(t)$ compared to one in Eq.~\eqref{fract51} and obtain
\beq
\frac{d\phi(t)}{ dt} = f_0 \eta(t) - (1 - \lam)\mu_0 \phi(t).\label{mfeqfinal}
\eeq

The implications of neglecting the term $f_0 \eta \phi(t)$ in Eq.~\eqref{fract51}, which generates multiplicative noise, are discussed in Sec.~\ref{sec:multnoise}. Here we note that the multiplicative noise term vanishes if the limit $N\rightarrow \infty$ is taken before the critical point is approached, that is, if the mean-field limit is taken before $\lambda\rightarrow 1$. However, for finite $N$ the situation is more subtle.

The starting point for the derivation of Eq.~\eqref{mfeqfinal} was Eq.~\eqref{richer2}, the mean-field equation for the richer chosen agent. If the chosen agent is poorer than the average of the other agents, similar arguments lead to the same equation as Eq.~\eqref{mfeqfinal}.

The form of Eq.~\eqref{mfeqfinal} is identical to the linearized version of the Landau-Ginzburg equation~\cite{ma, kl-batrouni, hh} with $\phi$
as the fluctuatng part of the order parameter. Hence, $\lam = 1$ corresponds to a phase transition as was found in simulations of the GED model~\cite{kang}. As for the usual Landau-Ginzburg equation, the factor of $(1 - \lam)$ sets the time scale for $\mu_0 \neq 0$. That is, as $\lam\rightarrow 1^-$, there is critical slowing down, and the system decorrelates on the time scale
\beq
\tau\sim \frac{1}{\mu_0(1 - \lam)}. \label{slowing}
\eeq

Because the stable fixed point of the poorer agent is zero for $\lambda > 1$ [see Eq.~\eqref{poorer}] and is one for the richer agent [see Eq.~\eqref{richer}], the order parameter is constant for both  $\lam > 1$ and $\lam < 1$, which indicates 
that there is a discontinuous jump in the order parameter at $\lambda = 1$. 
Hence, the exponent $\beta$, which characterizes the way the order parameter approaches its value at the transition, is equal to zero.

To obtain the critical exponent $\gamma$, we adopt an approach introduced by Parisi and Sourlas~\cite{ps} and note that the measure of a random Gaussian noise is given by~\cite{ps}
\begin{align}
P(\lbrace \eta_{j}\rbrace) & = \frac{\exp \Big[\!\int_{-\infty}^{\infty} -\beta \sum_{j} \eta_{j}^{2}(t)\,dt\Big]}{\!\int\! \prod_{j}\delta\eta_{j} \exp \Big[\!\int_{-\infty}^{\infty}-\beta \sum_{j} \eta_{j}^{2}(t)dt\Big]}, \label{par-sou} \\
\noalign{\noindent or}
P(\eta) & = \frac{\exp \Big[\!\int_{-\infty}^{\infty} -\beta N \eta^{2}(t)\,dt\Big]}{\!\int \delta\eta \exp \Big[\!\int_{-\infty}^{\infty}-\beta N \eta^{2}(t)dt\Big]}. \label{par-sou22}
\end{align}
The factor of $N$ in the argument of the exponential in Eq.~\eqref{par-sou22} comes from the fact that $\eta_{j}(t) = \eta(t)$ for all $j$ in the mean-field approach. This factor of $N$ is consistent with the argument that $\eta(t)\sim 1/{\sqrt N}$. 
(In Ref.~\cite{bigklein} the factor of $N$ is not explicit, but is implicit in the integral over all space.)

We rewrite Eq.~\eqref{mfeqfinal} as
\beq
\frac{1}{f_0}\frac{d\phi(t)}{dt} + \frac{(1 - \lam)}{f_0}\mu_0 \phi(t) = \eta(t), \label{mfeqfinal2}
\eeq
and replace $\eta(t)$ in Eq.~\eqref{par-sou22} by the left-hand side of Eq.~\eqref{mfeqfinal2}. This replacement requires a Jacobian, but in this mean-field case the Jacobian is unity~\cite{kl-batrouni}. 
Hence the, probability of $\phi$ is given by
\beq
P(\phi) = \frac{\exp \Big \{\!- \beta N \!\int_{-\infty}^{\infty} \Big [ \dfrac{1}{f_0}\dfrac{d\phi(t)}{dt} + \dfrac{\mu_0(1 - \lam)}{f_0}\phi(t)\Big ]^{2} \!dt\Big\} }{ \! \int\! \delta \phi(t)\exp \Big\{\!- \beta N \!\int_{-\infty}^{\infty} \Big [\dfrac{1}{f_0}\dfrac{d\phi(t)}{dt} + \dfrac{\mu_0(1 - \lam)}{f_0}\phi(t)\Big ]^{2} \!dt\Big \}}.\label{par-sou2}
\eeq

We now assume that the system is in a steady state so that $d\phi(t)/dt = 0$ over a time scale of the order of $1/\mu_0(1 - \lam)$.
Hence, the average $\lb \phi^2 \rb$ is given by
\begin{align}
\lb \phi^2 \rb & = \frac{\!\int \delta \phi\, \phi^{2}\exp \Big \{- \beta N \!\int_{-\infty}^{\infty}dt \Big [\dfrac{\mu_0(1 - \lam)}{f_0}\phi\Big ]^{2}\Big \} }{\!\int \delta \phi\, \exp \Big \{- \beta N \! \int_{-\infty}^{\infty}dt \Big [\dfrac {\mu_0(1 - \lam)}{f_0}\phi\Big ]^{2}\Big \}}\label{par-sou3} \\
& = \frac{\int \delta \phi\, \phi^{2}\exp \Big [- \beta N \dfrac{\mu_0(1 - \lam)}{f_0^2} \phi^{2}\Big] }{ \int \delta\phi\,\exp \bigg[-\beta N \dfrac{\mu_0(1 - \lam)}{f_0^{2}}\phi^{2}\bigg]}, \label{par-sou4}
\end{align}
where the range of integration over time is limited to the interval $1/\mu_0(1 - \lam)$.

Because we have assumed a steady state, the functional integral becomes a standard integral over $\phi$. We can take the limits of the integrals to be $\pm \infty$ because the factor of $N \gg 1$ in the exponential keeps $\phi$ of order $1/\sqrt N$. Hence, Eq.~\eqref{par-sou4} now becomes
\beq
\lb \phi^2 \rb = \frac{\int_{-\infty}^{\infty} d\phi\, \phi^2 \exp \Big \{\!- \beta N \dfrac{\mu_0(1 - \lam)}{f_0^2} \phi^2\! \Big \} }{ \int_{-\infty}^{\infty} d \phi\, \exp \Big \{-\beta N \dfrac{\mu_0(1 - \lam)}{f_0^2} \phi^{2}\! \Big \} }. \label{par-sou5}
\eeq
By using simple scaling arguments we see that the second moment of the probability distribution diverges as
\beq
\lb \phi^2 \rb \sim \frac{f_0^{2}}{N\mu_0(1- \lam)}. \label{eq:secondmoment}
\eeq

The fluctuating part of the order parameter $\phi = N\delta$ is analogous to the fluctuating part of the order parameter $m=M/N$ of the fully connected Ising model, where $M$ is the total magnetization of the system and $N$ is the number of spins. To determine the susceptibility (per spin) of the Ising model, we need to multiply $[\lb m^2 \rb - \lb m\rb^2]$ by $N$.
Because $\lb \phi^2 \rb =f_0^{2}[N\mu_0 (1 - \lam)]^{-1}$ [see Eq.~\eqref{eq:secondmoment}],
the susceptibility (per agent) of the GED model is given by
\beq
\chi \sim \frac{f_0^{2}}{\mu_0(1-\lam)}. \label{eq:chi}
\eeq
We conclude that the susceptibility diverges near the phase transition with the exponent $\gamma = 1$.

Note that we can relate the variance of $\phi$ to the variance of the rescaled wealth. From the definition of $\delta(t)$ in Eq.~\eqref{eq:delta}
and the fact that $x(t) = w(t)/W(t)$
is the rescaled wealth [see Eq.~\eqref{ydef}], we have
\beq
\phi(t) = 1 - Nx(t) = 1 - N\frac{w(t)}{W(t)} = 1 - N\wpp(t). \label{ndelta}
\eeq
We rescale the total wealth and hence the wealth of each agent so that $W(t) = N$ after the increased wealth due to economic growth has been assigned. Hence $\wpp$ in Eq.~\eqref{ndelta} is the rescaled wealth of a single agent.
Equation~\eqref{ndelta} will be useful in Sec.~\ref{sec:comparison} where we compare the predictions of the theory to the results of the simulations in Ref.~\onlinecite{kang}. 

\section{\label{sec:gedenergy}The Energy and Specific Heat Exponents}

From Eq.~\eqref{par-sou2} we have that
\beq
P(\phi) = \frac{\exp \Big \{\! - \beta N \mu_0 \dfrac{(1 - \lam)}{ f_0^{2}}\phi^{2}\Big\}}{\!\int d \phi\exp \Big \{\!- \beta N \mu_0 \dfrac{(1 - \lam)}{f_0^{2}} \phi^{2}\Big \}}, \label{par-sou7}
\eeq
assuming that the system is in a steady state. 
From the expression of the action or Hamiltonian in Eq.~\eqref{par-sou7}, where $\phi^2$ is multiplied by 
$\beta N\mu_0(1 - \lam)/f_0^{2}$, we see that the Ginzburg parameter for the GED model is given by (up to numerical factors) 
\beq
G = \frac{N\mu_0(1 - \lam)}{f_0^{2}}.
\label{eq:Ginzburg}
\eeq
To understand why $G$ on Eq.~\eqref{eq:Ginzburg} can be interpreted as the Ginzburg parameter, compare the form of Eq.~\eqref{par-sou7} with the form of the Hamiltonian for the \fc\ Ising model in Eq.~\eqref{H-fc-gauss} and the dependences of $G$ in Eqs.~\eqref{eq:Ginzburg} and \eqref{G-crit2} on their respective parameters.

The inverse temperature $\beta$ (not to be confused with the order parameter critical exponent), which arises from the amplitude of the Gaussian noise, will be absorbed in the parameter $f_0$. The association of $\beta$ with $f_0$ is consistent with Eq.~\eqref{mfeqfinal2} in that we are relating the temperature to the amplitude of the noise and indicates that increasing the 
fraction of the poorer agent's wealth transferred in an exchange is equivalent to increasing the amplitude of the noise.

The total energy for the GED model can be seen from the form of the action or the Hamiltonian in Eq.~\eqref{par-sou7}
\beq
E = N \phi^{2}, \label{eq:action}
\eeq
in analogy with the Landau-Ginzburg-Wilson free field or Gaussian action for the fully connected Ising model~\cite{lou}. Equations~\eqref{ndelta} and \eqref{eq:action} imply that the total energy of a system of $N$ agents is given by
\begin{subequations}
\begin{align}
E &= \sum_{i=1}^N(1-\wpp_i)^2\\
& = -N + \sum_{i=1}^N \wpp_i^2,\label{eq.eng}
\end{align}
\end{subequations}
where we have used that fact that $\sum_i \wpp_i = N$.

The existence of a quantity that can be interpreted as an energy implies that the probability density of the energy is given by the Boltzmann distribution for $\lam < 1$. The latter is consistent with simulations of the GED model~\cite{kang}. The existence of the Boltzmann distribution also implies that the system is in thermodynamic equilibrium and is not just in a steady state for $\lam < 1$.

From Eq.~\eqref{par-sou7} we find that $\lb \phi^2 \rb \sim f_0^2/[N\mu_0(1 - \lam)]$. Hence, we conclude from Eq.~\eqref{eq:action} that the mean energy per agent of the GED model scales as 
\beq
\frac{\lb E \rb}{N} \sim \frac{f_0^{2}}{N\mu_0(1-\lam)}. \label{av-eng-agent}
\eeq
Equation~\eqref{av-eng-agent} suggests that the mean energy per agent diverges as $(1-\lam)^{-1}$ as $\lam\rightarrow 1$ for fixed $N$, which is not physical. However, if we hold the Ginzburg parameter $G$ constant as $\lam \to 1$, we find no divergence (the exponent is zero), which removes the apparent nonphysical behavior. That is,
\beq
\frac{\lb E \rb}{N} \sim
\begin{cases}
(1-\lam)^{-1} & \mbox{(fixed $N$)} \\
G^{-1} & \mbox{(constant $G$)}.
\label{energy-both}
\end{cases}
\eeq
Equation~\eqref{energy-both} implies that the energy per agent is finite as we approach the critical point only if we hold $G$ constant. 

Near the critical point the nonanalytic behavior of the mean energy per agent can be expressed as $(1 - \lambda)^{1- \alpha}$, where $\alpha$ is the specific heat exponent. Equation~\eqref{energy-both} for $\lb E \rb/N$ for constant Ginzburg parameter implies that $\alpha = 1$. This result for $\alpha$ is what we would find if we require that $\beta$, $\gamma$, and $\alpha$ to satisfy the scaling relation in Eq.~\eqref{scalinglaw} with $\beta = 0$ and $\gamma=1$.

We can also calculate $\alpha$ directly using the probability distribution in Eq.~\eqref{par-sou7}. To calculate the fluctuations in the total energy, we need to calculate the average of
$\phi^{4}$. If we apply the probability in Eq.~\eqref{par-sou7}, we find that the fluctuations in the energy per agent, and hence the specific heat is proportional as $Nf^4[\mu_0 (1-\lambda)]^{-2}$,
where we have multiplied by $N$ as we did for the susceptibility per spin of the \fc\ Ising model. Hence, the specific heat $C$ scales as
\beq
C \sim \frac{f_0^4}{N \mu_0^2(1 - \lambda)^{2}}, \label{spec-heat}
\eeq
and 
\beq
\label{spec-heat2}
C \sim
\begin{cases}
(1-\lam)^{-2} & \mbox{(fixed $N$)} \\
(1-\lam)^{-1} & \mbox{(constant $G$)}.
\end{cases}
\eeq
We see that if we keep the Ginzburg parameter constant, we find $C \sim f_0^2/[G\mu_0(1 - \lambda)]$ and hence $\alpha = 1$. Note that if we do not keep $G$ constant, we would find $\alpha = 2$, which does not satisfy Eq.~\eqref{scalinglaw}.
As a consistency check, we can use Eqs.~\eqref{av-eng-agent} and \eqref{spec-heat} to construct the Ginzburg parameter by comparing the fluctuations of the energy, that is, the heat capacity, to the mean energy: 
\beq
\frac {NC}{ {\lb E\rb}^2} \propto \frac{f_0^2}{N\mu_0(1 - \lambda)} = G^{-1}.
\eeq

\section{\label{sec:comparison}Comparison with Simulations}

The mean-field theory predictions for the exponents $\alpha = 1$, $\beta=0$, and $\gamma=1$ are consistent with the simulation results reported in Ref.~\onlinecite{kang} for fixed $G$. As discussed in Sec.~\ref{sec:transition}, mean-field theory also predicts that there is only one time scale near the phase transition and that the time scale diverges as ($1-\lam)^{-1}$ for fixed Ginzburg parameter, an example of critical slowing down [see Eq.~\eqref{slowing}]. 
This prediction is consistent with the simulation results for the mixing time associated with the wealth metric~\cite{kang} and the energy decorrelation time, which were both found to diverge as $(1-\lam)^{-2}$ for fixed $G$. 
The apparent discrepancy between the $(1-\lam)^{-2}$ divergence found in the simulations and the $(1-\lam)^{-1}$ divergence predicted by Eq.~\eqref{slowing} is due to the difference in the choice of the unit of time in the simulation ($N$ exchanges) and in the mean-field theory ($N^2$ exchanges). To account for the difference in time units, we need to divide the simulation result by $N$ with the result that $N^{-1}(1-\lam)^{-2} \sim (1-\lam)(1-\lam)^{-2} = (1-\lam)^{-1}$, where we have used the relation $N \propto (1-\lam)^{-1}$ for fixed $G$ [see Eq.~\eqref{eq:Ginzburg}].

The simulations for fixed $G$ indicate that the energy per agent approaches a constant as 
$(1- \lambda)\rightarrow 0$. This behavior is associated with the nonanalytic part of the energy per agent. 
This result for the $\lambda$-independence of the nonanalytic part of the energy per agent is inconsistent with the relation between the energy per agent and the specific heat, $C \propto \partial \lb E (\lambda)\rb/\partial \lambda$. 
The $(1-\lambda)^{-1}$ dependence of the specific heat for fixed $G$ near $\lam=1$ suggests that the mean energy per agent could include a logarithmic dependence on $\lam$. For example, the form, $\lb E\rb/N \sim a_0 + a_L/\log(1-\lambda)$, where $a_0$ and $a_L$ are independent of $\lam$, implies that the specific heat scales as $C \sim [\log(1-\lam)]^{-2}(1-\lam)^{-1}$, thus yielding $\alpha=1$ with logarithmic corrections, which standard mean-field theory cannot predict and are very difficult to detect in simulations.

There is also agreement between the exponents predicted by \mf\ theory and those determined in the simulations 
when the measurements are done at fixed $N$. 
From Eq.~\eqref{av-eng-agent} we see that if $N$ is held constant, the mean energy per agent is predicted to diverge as $(1-\lam)^{-1}$, which is consistent with the simulations~\cite{kang}, although this divergence, is unphysical because it implies that the mean energy per agent would become infinite. The exponent $\alpha$ is predicted to be equal to 2 for fixed $N$, which is also in agreement with the simulations~\cite{kang}.

\section{\label{sec:multnoise}Multiplicative Noise}

% fig1
\begin{figure}[t]
\includegraphics[scale=0.6]{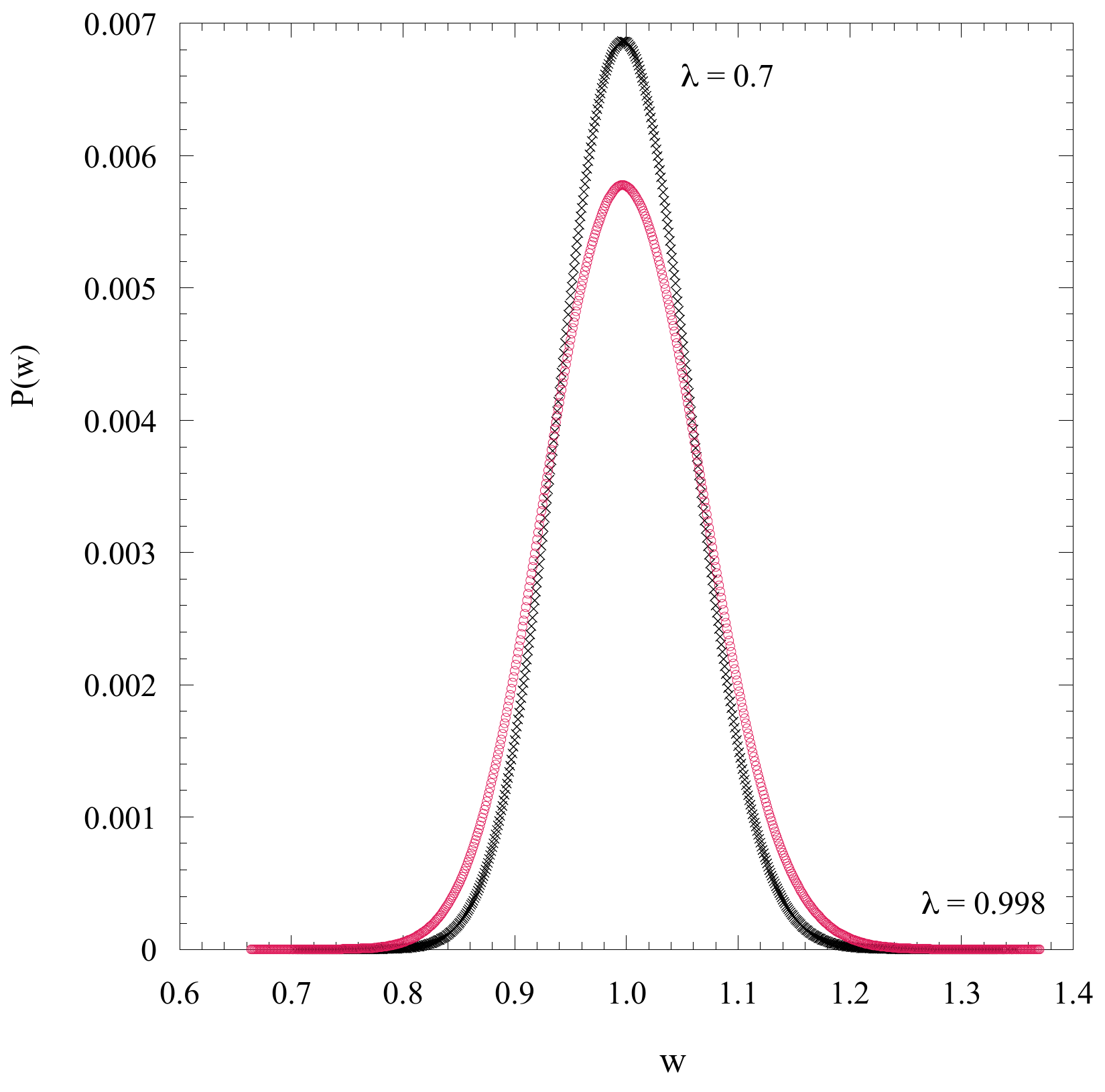}
\vspace{-0.2in}
\caption{\label{fig:wealthhdist}Comparison of the wealth distribution $P(w)$ for $\lambda=0.998$, $N=5 \times 10^5$, and $M\approx 14$ (red curve) with $P(w)$ for $\lam = 0.700$, $N=3333$, and $M \approx 173$ (more sharply peaked black points). Both plots are for $G=10^6$, with $f_0=0.01$, and $\mu_0=0.1$. The two distributions would be identical if mean-field theory were exact. Plots of $P(w)$ for closer values of $M$ [see Eq.~\eqref{eq:M}] are indistinguishable to the eye. Values of $\lam = 0.7$ and 0.998 were chosen so that the differences of $P(w)$ are noticeable in the plot.)}
\end{figure}

A sensitive test of whether the system is in equilibrium 
is given by the form of the wealth distribution of the agents. The derivation of the Gaussian form of the wealth distribution [see Eq.~\eqref{par-sou7}] assumes that the system is in a steady state and that $G \to \infty$ and implies that the distribution of the energy is a Boltzmann distribution. 
The wealth distribution $P(w)$ in Eq.~\eqref{par-sou7} is predicted to depend only on the value of $G$ and not on the parameters $\lambda$, $f_0$, and $\mu_0$ separately. 
Figure~\ref{fig:wealthhdist} shows the distribution of wealth for fixed $G=10^6$ and different values of $\lambda$ and $N$.  Although the distributions are similar, we see that the wealth distribution is not invariant with respect to changes of $\lambda$ for fixed $G$, even though both distributions are well fit by a Gaussian. Similar changes in $P(w)$ are found for changes in the other parameters for fixed $G$.

To understand this behavior, we return to Eq.~\eqref{fract51}, the mean-field equation for the evolution of the wealth near the $1/N$ fixed point, which we repeat here for convenience:
\beq
\frac{d\phi(t)}{dt} = f_0 \eta(t) \big [1 - \phi(t)\big] - \mu_0 (1 - \lam)\phi(t). \tag{\ref{fract51}}
\eeq
In Sec.~\ref{sec:transition} we argued that the multiplicative noise term $f_0 \eta \phi$ can be neglected because $\phi$ is assumed to be much less than one. However, we retained the ``driving'' term $\mu_0(1 - \lam)\phi$ and did not consider whether the multiplicative noise term was small compared to the driving term. To determine if this condition holds, we recall that the (average) noise $\eta$ is assumed to be random Gaussian. Our assumption that the amplitude of the Gaussian noise is proportional to $1/\sqrt N$ is consistent with the dependence of the Gaussian noise in the mean-field limit of thermal models such as the Ising model (see, for example, Ref.~\onlinecite{bigklein}).

Because the amplitude of the Gaussian noise is of order $1/\sqrt N$, we can neglect it compared to the driving term in Eq.~\eqref{fract51} if $f_0/\sqrt N \ll \mu_0(1-\lam)$, or
\beq
M \equiv \frac{{\sqrt N}\mu_0(1 - \lambda)}{f_0} \gg 1. \label{eq:M}
\eeq
Equation~\eqref{eq:M} defines the parameter $M$. The condition $M \gg 1$ for the neglect of the multiplicative noise term, as well as the condition $G \gg 1$ has several implications.

\begin{enumerate}
\item Both $G$ and $M$
diverge in the mean-field limit for which first $N \to \infty$ and then the critical point at $\lambda = 1$ is approached~\cite{kac}. If these limits are taken in this order, the mean-field treatment neglecting the multiplicative noise is exact 
(see Ref.~\cite{bigklein} and references therein).

\item A smaller value of $f_0$ makes the system more describable by a mean-field treatment, which explains the better agreement of the exponents determined from the simulations for finite values of $N$ with the exponents calculated from a theory that neglects the multiplicative noise.

\item A large value of $G$ does not necessarily imply a large value of $M$; that is, as $\lam \to 1$, the multiplicative noise can become important even though $G$ is still much greater than one.

\item The Ginzburg parameter $G$ controls the level of mean field and $M$
controls the influence of the multiplicative noise. It is necessary to keep both parameters constant to obtain results consistent with the \mf\ theory. Because we cannot keep both parameters constant simultaneously, there will always be some inconsistency of the results for finite values of $N$ and $G$. These inconsistencies can be minimized for sufficiently large $N$ by increasing $\mu_0$ or decreasing $f_0$. 
The point is that we need to be careful in interpreting the results of simulations. An example of the limitations of the \mf\ theory and the neglect of both the additive and multiplicative noise terms is shown in Table~\ref{tab:dependence}. We see that both $\tau_E$, the energy decorrelation time, and $\tau_m$, the mixing time, depend weakly on $f$ in contrast to Eq.~\eqref{slowing} which predicts that these times are independent of $f$. The dependence of $\tau_E$ and $\tau_m$ on $f$ reflects the possible importance of the multiplicative noise.
\end{enumerate}

\begin{table}[t]
\begin{tabular}{|l|l|l|r|r|}
\hline
$N$ & $f$ & $\mu$ & \hfill $\tau_m$ \hfill & $\tau_E$ \\
\hline
5000 & 0.01	& 0.01 & 1034 &	229 \\

5000 & 0.10 &	0.01 & 1229 &	149 \\
5000 & 0.10 & 0.10 & 116	& 21 \\ 
1000 & 0.10 & 0.10 & 115 &	21 \\
\hline
\end{tabular}
\caption{\label{tab:dependence}Summary of the dependence of the mixing time $\tau_m$ and the energy decorrelation time $\tau_E$ on $N$, $f$, and $\mu$ for $\lam=0.8$ ($f$ and $\mu$ are not scaled). Comparison of the first two rows indicates that $\tau_m$ and $\tau_E$ depend weakly on $f$ for fixed $N$ and $\mu$. Comparison of the second and third rows suggests that $\tau_m$ and $\tau_E$ depend strongly on the value of $\mu$. Comparison of the third and fourth rows indicates that $\tau_m$ and $\tau_E$ are independent of $N$. These dependencies are in qualitative agreement with Eq.~\eqref{slowing}.
}
\end{table}

\section{\label{sec:geometric}Relation to the Geometric Random Walk}

For either zero growth, $\mu_0 = 0$, or for the critical point, $\lam = 1$, the mean-field equation for the rescaled wealth, Eq.~\eqref{poorer}, reduces to
\beq
\frac{dx(t)}{dt} = f_0\eta(t) x(t). \label{eq:mfeqq}
\eeq
If we use the Ito interpretation for the effect of the multiplicative noise in Eq.~\eqref{eq:mfeqq}, the solution for $x(t)$ is
\beq
x(t) = x(t=0)\exp\Big [-\frac{f_0^{2} }{2}t + f_0 W_{t}\Big ],\label{grw}
\eeq
where $W_{t}$ is a Brownian noise or Wiener process and is given by
\beq
W_{t} = \! \int^{t}\! \eta(t')dt'. \label{brown}
\eeq
Because Eq.~\eqref{eq:mfeqq} results from either setting $\lam = 1$ or $\mu_0 = 0$, Eq.~\eqref{eq:mfeqq} implies that the mean-field treatment of the GED model for $\mu_0 = 0$ and $\lam \neq 1$ results in the same distribution as the geometric random walk without the drift term~\cite{peters,peters-klein}. For $\lam = 1$ and $\mu_0 \neq 0$, the solution is $e^{\mu_0 t}x(t)$, where $x(t)$ is the solution with $\mu_0=0$, and Eq.~\eqref{grw} describes the distribution of the geometric random walk with the drift or growth term~\cite{peters,peters-klein}.

This result, which follows from the analysis of the mean-field equation, Eq.~\eqref{poorer}, is not applicable if $N$ is held constant because $G = 0$ for $\mu_0=0$ or $\lam=1$, and hence the mean-field approach does not apply. If we keep $G$ constant, Eq.~\eqref{eq:mfeqq} is applicable because the condition $G \gg 1$ is compatible with either $\mu_0 \approx 0$ or $\lam \approx 1^-$. To show numerically that the GED model reduces to the geometric random walk at the critical point involves fixing the value of $G$ and determining the form of the wealth distribution for $\lam \neq 1$ and $\mu_0>0$ and then extrapolating the wealth distribution in the limit $\lam \to 1$ or $\mu_0 \to 0$. Such an extrapolation would be a difficult and time consuming process.

\section{\label{sec:discussion}Summary and Discussion}

We have investigated a simple agent-based model of the economy in which two agents are chosen at random to exchange a fraction of the poorer agent's wealth. Economic growth is distributed 
according to the parameter $\lam$. 
The larger the value of $\lam$, the greater the fraction of the growth that is distributed to the agents at the higher end of the wealth distribution. The model, which we call the GED model, was treated theoretically with a mean-field approach and was shown to have a critical point at $\lam = 1$, consistent with simulations~\cite{kang}. 
The critical exponents are consistent with scaling and the simulations if the Ginzburg parameter is large and held constant as the critical point is approached.

The agreement of the mean-field theory with the simulations implies that for finite but large $G$ and $M$, the GED model can be characterized as near-\mf~\cite{bigklein}. That is, the system is \mf\ in the limit $N\rightarrow \infty$, and is well approximated by mean-field theory if both $N$ and $M \gg 1$, provided that the Ginzburg parameter $G \gg 1$ and is held constant as the transition is approached.

The mean-field theory and the simulations raise some interesting questions about the relation between growth, uncertainty and wealth inequality and the applicability of 
statistical physics. The questions concerning statistical physics include the following:

\begin{itemize}

\item The inclusion of distribution and growth allows the system to be treated by the methods of equilibrium statistical mechanics,
but only if the distribution parameter $\lambda < 1$ and in the limit that the number of agents $N\rightarrow \infty$. A similar result holds for models of
earthquake faults for long-range stress transfer~\cite{bigklein, rundle}. It is unclear how many driven dissipative 
non-equilibrium systems become describable by equilibrium methods in the mean-field limit. 

\item We used equilibrium methods to calculate the critical exponents in agreement with the simulations, but the exponent $\alpha$ associated with the specific heat is thermodynamically consistent only
if the Ginsburg parameter is held constant. 
Similar results were found for the fully connected Ising model~\cite{lou, kangspecificheat}. 
Insight into why holding $G$ constant is necessary will be discussed in detail in a future publication~\cite{kangspecificheat}.

\item A subtle feature of using a \mf\ approach to treat the GED model for $N\gg 1$ but finite is the presence of multiplicative as well as additive noise. The effect of the multiplicative noise is controlled by the parameter $M$ defined in Eq.~\eqref{eq:M}. 
From the agreement of the theory with the simulations, we conclude that the neglect of the multiplicative noise in the theory is a good approximation for $M\gg 1$. 
The role of multiplicative noise is of particular interest for models of the economy in light of the non-ergodicity of the geometric random walk, which includes multiplicative noise~\cite{peters,peters-klein}. 
 
\item To obtain an equilibrium description of critical point behavior, we defined an order parameter and then obtained the order parameter exponent $\beta$ and the susceptibility exponent $\gamma$. To obtain the 
specific heat exponent $\alpha$, we defined an energy, which also allowed us to obtain the $\lam$ dependence of the energy as $\lambda$ approaches its critical value. The definitions of the order parameter and the energy generate a thermodynamically consistent set of  exponents that characterize the critical point. Is our choice of order parameter and energy unique, or are there other definitions that would lead to another set of thermodynamically consistent exponents?
 
\begin{comment}
[xx omitted because we mention networks in the last paragraph and becaue we don't seem to be able to determine $\nu$.
\item We have considered a version of the GED model that has no geometry; that is, an agent can exchange wealth with any other agent. An interesting question is whether our results are altered by including a geometry. We are pursuing this question by simulating the model on a regular lattice in one and two dimensions as well as on scale free and Erd\"os-Renyi networks. 
Including a geometry, particularly for regular lattices, will allow us to define a correlation length and investigate 
the correlation length exponent $\nu$ and the application of hyperscaing ideas and their relation to a fixed Ginzburg parameter~\cite{bigklein}. 
\end{comment}

\end{itemize}

Any statements about a system as complicated as the economy based on the simple GED model must be viewed with a considerable amount of caution. However, the results obtained from both the numerical and theoretical investigations of the GED model suggest some general properties of economic systems that are of potential interest.

\begin{itemize}

\item The form of the exchange term in Eq.~\eqref{exacteq} assumes that the amount of the exchange is determined by the poorer agent. This assumption is a reasonable first approximation because in most exchanges of goods or services, the poorer of the two agents decides if they can afford the exchange. The fact that the wealth transferred is a percentage of the poorer agent's wealth leads to the multiplicative part of the noise. 

\item The exchange term in Eq.~\eqref{exacteq} also assumes that the winner of the exchange is based on the toss of a true coin. Such a toss assumes that both agents have equal knowledge of the worth of the exchange at the time of the exchange, so that any advantage enjoyed by the winning agent is gained by pure chance.
The effect of biasing the coin toss to represent a superior knowledge of either the richer or poorer agent is a subject of future study~\cite{Cardoso}.

\item We found that if the distribution of the wealth generated by economic growth is not skewed too heavily toward the wealthy ($\lambda < 1$), then every agent's wealth grows exponentially with time. The distribution of wealth is not equal, but wealth condensation is avoided. As $\lam \to 1^-$, the wealth distribution becomes more skewed toward the wealthy, thus increasing inequality. The theory indicates that a more unequal distribution of added wealth due to growth can be overcome by either increasing the growth parameter $\mu_0$, decreasing the uncertainty by decreasing $f_0$, or by increasing $N$. The theory also indicates that there is a tipping point at $\lam = 1$, so that for $\lam \geq 1$, no increase of $\mu_0$ or decrease in $f_{0}$ can overcome the inequality caused by the distribution of the growth favoring the wealthy. Although the GED model is very simple, this result raises the question of whether there is a tipping point in more realistic models of the economy. That is, can the distribution of the growth in wealth favor the rich to such an extent that the
increased wealth (``a rising tide'') is no longer shared by the majority of people (``lifts all boats''), and the effect of the unequal added wealth distribution cannot be 
alleviated by increased growth or decreased uncertainty?

\item The theory suggests that as the number of agents $N$ is increased, with the parameters $\lam$, $f_{0}$, and $\mu_{0}$ held fixed, the system becomes more describable by a mean-field approach. 
This result suggests that as globalization increases, mean-field models of the global economy might become more relevant and equilibrium methods might be more appropriate in contrast to economic models that are not ergodic~\cite{peters, peters-klein, petersnature}. We stress that an equilibrium treatment would be an approximation and be exact only for $N\rightarrow \infty$, but might be a good approximation for $N \gg 1$, assuming that $G$ and $M$ are both much greater than one. The question of how the multiplicative noise would affect the 
system if simulated for a very long time is not clear. 
We found that if the effect of the multiplicative noise is increased by lowering $M$, the wealth distribution develops a tail for large wealth, indicating that the multiplicative noise induces greater wealth inequality.

\item The model also suggests that increasing the noise amplitude $f_0$ increases wealth inequality. In addition, the theory assumes that the parameters $\lambda$, $f_{0}$, and $\mu_{0}$ are independent. 
These parameters are not necessarily independent in actual economies, which
raises the question of how these variables affect each other. For example, $\mu_{0}$ could be made to depend on $\lambda_{0}$. If $\mu_0$ is increased as $\lambda_{0}$ is increased, this dependence would be a test (in the model) of the trickle down theory. 

\end{itemize}

Besides the areas of future research raised by these questions, 
other areas include investigating the effect of growth in models on various network topologies 
and investigating different exchange mechanisms and how they affect the distribution of wealth when growth is added. 

\appendix

\section{\label{sec:Appendix}Appendix: The Fully Connected Ising Model}

It is useful to discuss the analogous equilibrium behavior of the \fc\ Ising model. To do so, we first consider the long-range Ising model with interaction range $R$ in the limit that $R \to \infty$.

The Landau-Ginzburg-Wilson Hamiltonian for the long-range Ising model in zero magnetic field is given by~\cite{bigklein}
\beq
\label{LGFCIM}
H\big (\phi({\vec y})\big) = \! \int\! d{\vec y}\, \Big [R^{2}(\nabla \phi({\vec y}))^{2} + \epsilon\phi^{2}({\vec y}) + \phi^{4}({\vec y})\Big ].
\eeq
The integral in Eq.~\eqref{LGFCIM} is over all space, $\epsilon = (T - T_{c})/T_{c}$, $T_{c}$ is the critical temperature, and $\phi(\vec y)$ is the coarse grained magnetization. 

Near the mean-field critical point we scale $\phi(\vec y)$ by $\epsilon^{1/2}$, scale all lengths by 
$R\epsilon^{-1/2}$, and obtain
\beq
\label{LGFCIMsc}
H\big ((\psi({\vec x})\big ) = R^{d}\epsilon^{2 - d/2}\! \int d{\vec x}\, \Big [ (\nabla\psi({\vec x})^{2} + \psi^{2}({\vec x}) + \psi^{4}({\vec x})\Big ],
\eeq
where $\psi({\vec x}) = \epsilon^{-1/2}\phi(\vec y/R)$ and ${\vec x} = \vec y/R\epsilon^{-1/2}$. The integral is over the volume in scaled coordinates. Because the functional integral over $\psi({\vec x})$ is damped for larger values of $\psi$ due to the Boltzmann factor $e^{-\beta H(\psi({\vec x}))}$ and $R^{d}\epsilon^{2-d/2}\gg 1$, the rescaled magnetization $\psi({\vec x})$ satisfies the condition, 
\beq
\label{psi-ineq}
\psi({\vec x})< {\sqrt \frac{1}{R^{d}\epsilon^{2 - d/2}}}.
\eeq
For the fully connected Ising model
we can ignore the gradient term in $H$ and take $R^{d}\epsilon^{2 - d/2}\rightarrow N\epsilon^{2}$,
and the integral in Eq.~\eqref{LGFCIMsc} becomes of order one. 

To calculate the exponent $\beta$ for the fully connected Ising model, we take $\epsilon < 0$ and write
\beq
\label{LGW-beta}
H\big (\psi \big) = N\epsilon^{2} \big [- \psi^{2} + \psi^{4}\big ].
\eeq
The most probable value of $\psi$ is obtained by setting the derivative with respect to $\psi$ of $H(\psi)$ equal to zero; the result is 
that the most probable value of $\psi$ is $\sim \epsilon^{1/2}$ and $\beta = 1/2$.

To calculate the isothermal susceptibility $\chi$ for the \fc\ Ising model, we can ignore the quadratic term in Eq.~\eqref{LGW-beta} and write the action of Hamiltonian as
\beq
\label{H-fc-gauss}
H(\psi) = N\epsilon^{2}\psi^{2}.
\eeq
We determine the probability as a function of $\psi$ and multiply the average of $\epsilon\psi^{2}$ ($\phi^2$) 
by $N$ to obtain $\chi \sim \epsilon^{-1}$ as expected.

Note that the action in Eq.~\eqref{H-fc-gauss} is order one for the range of fluctuations in the \fc\ Ising model; that is, $\psi\lesssim 1/{\sqrt {N\epsilon^{2}}}$. Similarly, we expect the action in the GED model to also be of order one, not order $N$.

The energy per spin of the fully connected Ising model is the square of the magnetization per spin~\cite{bigklein}. Hence the 
mean energy per spin is the average of $\epsilon\psi^{2} = 1/N\epsilon$. 
This dependence on $\epsilon$ seems nonphysical and seems to imply that the energy per spin diverges as $\epsilon \rightarrow 0$. To understand this result and to calculate the specific heat, we introduce the Ginzburg criterion, which is a self-consistency check on the applicability of mean-field theory~\cite{bigklein}. For a mean-field 
theory to be a good description, the fluctuations of the order parameter must be small compared to the mean value of the order parameter. This requirement implies that
\beq
\frac{\xi^{d}\chi}{\xi^{2d} \phi^{2}} = \frac1G \ll 1, \label{G-crit2}
\eeq
where $\xi$ is the correlation length, $\chi$ is the susceptibility, and $d$ is the spatial dimension. The Ginzburg parameter $G$ defined by Eq.~\eqref{G-crit2} must be much greater than one for mean-field theory to be a good approximation. Much numerical and theoretical work has shown that the Ginzburg criterion is a good indicator of the appropriateness of a mean-field description~\cite{ma, bigklein}. It is in this sense that we will use the Ginzburg criterion in the following.

Equation~\eqref{G-crit2} implies that the Ginzburg parameter for the fully connected Ising model is given by $G = N \epsilon^2$ (up to numerical factors). Mean-field theory for the fully connected Ising model becomes exact if the limit $N\rightarrow \infty$ is taken before $\epsilon \to 0$~\cite{bigklein}. As $\epsilon$ decreases for fixed $N$, $G$ decreases, which implies that the system becomes less describable by mean-field theory. To determine the critical exponents for the fully connected Ising model for a large but finite value of $N$ in a simulation, we need to keep the system at the same level of mean field, which implies that we must keep $G$ constant. Hence, as $\epsilon \to 0$, we need to consider larger and larger values of $N$. Keeping $G$ constant has the additional consequence of restoring two exponent scaling, which is missing in the standard treatments of mean-field systems~\cite{tane, lou}.

Another conclusion that follows from the Ginzburg criterion is that the scaling of the isothermal susceptibility $\chi$ must be the same as the scaling of $\xi^{d}\phi^{2}$ or $N\phi^{2}$ in the fully connected Ising model, which justifies multiplying the square of the average of $\phi = \epsilon^{1/2}\psi$ by $N$ to obtain $\chi$. 

Because we need to hold $G = N\epsilon^{2}$ constant to find consistent results for the mean-field Ising exponents, the result that the mean energy per spin is proportional to $1/N\epsilon$ can now be properly interpreted. We have
\beq
\label{mean-eng-G}
\frac{\lb E \rb}{N} = \frac{1}{N\epsilon} = \frac{\epsilon}{ N\epsilon^{2}} = \frac{\epsilon}{G} \sim \epsilon,
\eeq
where we have assumed that $G$ is a constant. This result is what is expected from a mean-field calculation because the nonanalytic part of the mean energy per spin should scale as $\epsilon^{1 - \alpha}$, with the mean-field value of the specific heat exponent 
$\alpha = 0$. 

We next calculate the specific heat of the \fc\ Ising model by recasting the Ginzburg criterion in terms of the energy fluctuations. For mean-field theory to be applicable, the fluctuations of the energy must be small compared to the square of the mean energy, or
\beq
\label{G-energy}
\frac{\xi^{d}C}{\xi^{2d}e^{2}} = \frac{C}{N\epsilon^{2}} = \frac{C}{G} \rightarrow C,
\eeq 
where $C$ is the specific heat. As expected, holding $G$ constant implies that the specific heat exponent $\alpha = 0$.
If we hold $N$ rather than $G$ constant, we would obtain $1 -\alpha = -1$ and $\alpha = -2$. We see that the two results for $\alpha$ are not consistent unless $G$ is held constant.

Note that the exponents $\beta = 1/2$ and $\gamma = 1$ are the same whether we hold $N$ or $G$ constant, but the value of $\alpha$ depends on whether $N$ or $G$ is held constant. Also the scaling relation \eqref{scalinglaw} cannot be satisfied for $\gamma = 1$ and $\beta = 1/2$ unless $\alpha = 0$ which in turn implies that we need to keep $G$ constant (and large) to obtain a consistent mean-field description.

Also note that the form of the right-hand side of Eq.~\eqref{H-fc-gauss} is the same as the action or Hamiltonian that we derived for the GED model using the Parisi-Sourlas method with $\epsilon^{2}$ replaced by $\mu_0(1 - \lambda)/f_0^{2}$ [see Eq.~\eqref{eq:action}]. 

\begin{acknowledgments}

We thank Ole Peters, Jon Machta, Jan Tobochnik, Bruce Boghosian, and Alex Adamou for useful discussions. 
WK would like to acknowledge the hospitality of the London Mathematical Laboratory where part of this work was done.

\end{acknowledgments}

\end{document}